# Lattice Energy Reservoir in Metal Halide Perovskites

Xiaoming Wen and Baohua Jia
RMIT University, Melbourne, 3000, Australia
E-mail: xiaoming.wen@rmit.edu.au, baohua.jia@rmit.edu.au

## Abstract

Metal halide perovskite-based technologies have been rapidly developed during the last decade. However, to date, the fundamental question "Why are halide perovskites superior to conventional semiconductors?" has remained elusive. Here, we propose a new theory of lattice energy reservoir (LER) in halide perovskites and elucidate that LER can comprehensively impact charge carrier dynamics and thus enhance device performance, from hot carrier cooling,[1-2] carrier recombinations,[3-4] anomalous upconversion fluorescence,[5-6] illumination induced fluorescence enhancement (photobrightening)[7-9], to high efficiency solar cells and light-emitting diodes. An LER is a dynamic nanodomain in halide perovskites with suppressed thermal transport that can accumulate energy from phonon coupling and then feedback to subgap carriers and result in subgap carrier upconversion. The LER directly results in slowed cooling of hot carriers and significantly prolonged carrier recombination, anomalous upconversion fluorescence, as usually termed as defect tolerance,[7, 10] as well as the anomalous ultraslow phenomena including persistent polarization,[11] memory effect,[12-13] and photobrightening. The LER theory rationalizes the superior optoelectronic properties and device performance and provides a novel physical understanding for anomalous phenomena observed uniquely in halide perovskites.

**Key words:** lattice energy reservoir, halide perovskites, dynamic defect tolerance, prolonged carrier lifetime, upconversion fluorescence

## 1. Introduction

Halide perovskites have emerged as a class of materials with unparalleled potential for revolutionizing optoelectronic devices, ranging from solar cells and light-emitting diodes to photodetectors and lasers. The exceptionally high performances have been commonly attributed to their unique optoelectronic properties, such as long carrier lifetime, long diffusion length, small binding energy, and defect tolerance.[14-15] Despite intensive effort towards understanding the underlying physics, the fundamental physics underlying the superior optoelectronic properties of halide perovskites remains elusive. In addition, many anomalous phenomena observed in halide perovskites challenge our current semiconductor theory, such as an exceptionally longer carrier recombination lifetime than that predicted by Langevin theory,[16-17] slowed cooling of hot carriers,[4] defect tolerance[7, 10, 18], anomalous upconversion fluorescence (exclusive multi-photon absorption and lanthanum-doping)[6, 19], as well as persistent structural polarization[11], memory[13, 20] and ultraslow variations in photoluminescence (PL) efficiency and carrier lifetime (also referred as to defect healing, defect curing, photobrightening)[9, 21-24]. At the heart of this challenge lies the absence of a unified theoretical framework capable of capturing the complex physics governing halide perovskites. To date, the understanding of these anomalous phenomena mostly stays at the phenomenological explanation level and lacks detailed physical mechanisms. The most fundamental question remains elusive: What makes halide perovskites superior over conventional semiconductors? What is the fundamental cause of their superior optoelectronic properties and anomalous phenomena in devices? In this context, we elucidate that lattice energy reservoir (LER) can comprehensively impact the carrier dynamics, from slowed cooling of hot carriers, charge carrier recombinations, to photobrightening and anomalous upconversion fluorescence. We therefore propose a new theory that LER in halide perovskites is the fundamental origin of the exceptional optoelectronic properties, to interpret and rationalize the diverse range of anomalous phenomena observed in halide perovskites.

## 2 Concept of lattice energy reservoir

Halide perovskites have been confirmed to exhibit unique a soft lattice.[11, 25] There is pronounced dynamical lattice distortion that substantially alters the electronic band structure.[26-27] The LERs are defined as the dynamic nanodomains ununiformly localized in the lattice of halide perovskites.[28] The prerequisite for a LER is extremely low energy dissipation (suppressed thermal transport at the interface). Upon illumination or other forms of phonon energy injection, the local lattice of LER can gradually distort and form homojunction with different lattices between internal and external LER with different potential energies. The strain at the interface homojunction can suppress thermal transport, which warrants that phonon energy can be efficiently stored in the LER.

A hot LER can be established through phonon-lattice coupling[29-31], as schematically shown in Figure 1, local lattice of LER may occur through contraction/expansion, deformation, and/or distortion. The soft lattice and sublattice of halide perovskites can interact with photogenerated electrons as polarons and exhibit various physical effects.[32-33] Using terahertz (THz) spectroscopy, Yue et al. observed the strong coupling between the carrier and lattice as a polaron. Anharmonic polaron resonances of inorganic sublattice I-Pb-I vibration mode and organic cation FA+ rotation mode are identified at 1 THz and 0.4 THz, respectively.[34] Then the LER will have a different lattice from the surround, and thus a special homojunction will form with an interface strain. This homojunction can significantly suppress thermal transport (decreased energy dissipation rate), and a higher potential energy could be gradually established in the LER higher over the surrounding. Consequently, the perovskites with such LERs will exhibit macroscopically ultralow thermal conductivity[35-37] and an ultralong dissipation lifetime. Therefore, the dynamic nanodomains will be non-uniformly localized in the perovskite lattice under continuous illumination. As discussed in detail later, the hot LERs will comprehensively impact carrier dynamics and eventually result in superior optoelectronic properties and exceptional device performance.

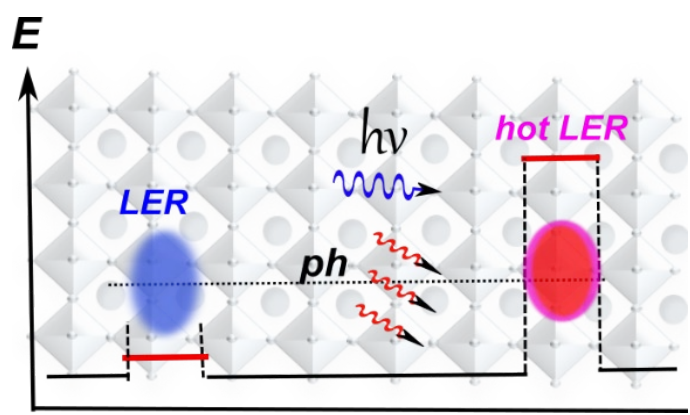

Figure 1. Scheme of Lattice energy reservoir as dynamic nanodomain in halide perovskites. Blue LER: before energy storage, red LER: energy accumulation by phonon-lattice coupling, higher potential energy is established over the surround.

The LER hypothesis has been supported by various observations in halide perovskites. Zhao *et al*. unveiled that there exists pronounced dynamical lattice distortion in the form of disordered instantaneous octahedral tilting and the lattice distortion substantially alters the electronic band structure through renormalizing the band dispersions and the interband transition energies.[27] Combined multiple microscopy experiments and molecular dynamic simulation, Dubajic and coworkers uncovered the existence of dynamic, lower-symmetry local nanodomains embedded within the higher-symmetry average phase in various perovskite compositions.[28, 38] The dynamic nanodomains are directly linked to the differing macroscopic optoelectronic and ferroelastic behaviors. Gao *et al*. observed atomically thin halide perovskite nanowires and confirmed the unusual dynamical behaviors pertaining to the lattice anharmonicity.[26] Using high angle annular dark-field transmission electron microscopy (HAADF-TEM), Cai and coworkers observed atomic-scale structures of intragrain stacking-fault and twinning interface, showed in-plane strain distribution in a single stacking fault and a dramatic lattice distortion at the stacking fault region.[39] Recently, Yang and coworkers demonstrated an inhomogeneous strain gradient at the interface leads to a drastically suppressed thermal conductivity.[35] It has been confirmed efficient optical to acoustic phonon upconversion occurs in halide perovskites due to suppressed thermal dissipation, which displays as extremely low thermal conductivity in macroscopy.[1, 37, 40] It is reasonable to suppose that the local homojunction at LER surface can efficiently suppress thermal dissipation; therefore the LER with lattice strain on the surface can store the phonon energy due to the ultralow thermal dissipation rate. The establishment of hot LER can be a slow accumulation process, that is, the LER density can be dependent on the excitation conditions, for example, light illumination intensity and wavelength, accumulation time, which intimately impact the observed phenomena.

## 3. Physical effects of Hot Lattice Energy Reservoirs on carrier Dynamics

The presence of hot LERs exerts a profound influence on the photogenerated carrier dynamics of halide perovskites, giving rise to a plethora of intriguing phenomena, including slowed carrier cooling, prolonged carrier recombination and mitigated nonradiative recombination by upconverting subgap carriers, ultimately leading to dynamic defect tolerance for exceptional device efficiency, and the emergence of anomalous ultraslow phenomena.

### 3.1 Slowed cooling of hot carriers

The hot carrier cooling is accompanied by hot phonon emission of longitudinal-optical (LO) phonons, which then decay to low-energy longitudinal-acoustical (LA) phonons, followed by an effective thermal transport to the surroundings via acoustic phonon propagation and lattice heating. Through this process, the excess kinetic energy of hot carriers is eventually dissipated into the surrounding lattice in an irreversible manner. Three distinct relaxation stages include: (1) hot carrier thermalization (Fröhlich interaction) via carrier-carrier, the hot carriers reach a quasi-equilibrium with the Fermi-Dirac distribution in the timescale of 100 fs. The temperature of the hot carrier system can be acquired from its transient distribution. (2) Hot LO phonons decay to LA phonons by the symmetric Klemens-decay and anti-symmetric Ridley-decay[41]; where one LO phonon decays into two LA phonons with energy and momentum conservation. (3) Acoustic phonons propagate and dissipate their energy into the surrounding lattice. An efficient acoustic phonon upconversion can recycle thermal energy back and reheat carriers, Figure 2a.[28]

Up to hundreds of picoseconds of slowed cooling have been confirmed in hybrid halide perovskites at high excitation density[1, 42-43] and the cooling time exhibits apparently excitation density dependence, that is, the cooling time is still short, few picoseconds, at low carrier density. At high density excitation, a high density of hot carriers can gradually establish a high density of hot LERs because the LER homojunction interface can impede thermal depletion in the surrounding lattice. As a result, high density acoustic phonon can accumulate so that the relaxation channel of hot carriers, from optical phonons to acoustic phonons then dissipating into the environment, is suppressed. The slowed cooling can be observed due to acoustic-optical phonon upconversion[1] or Auger heating[2], which results in the slowed cooling of hot carriers.[1, 43]

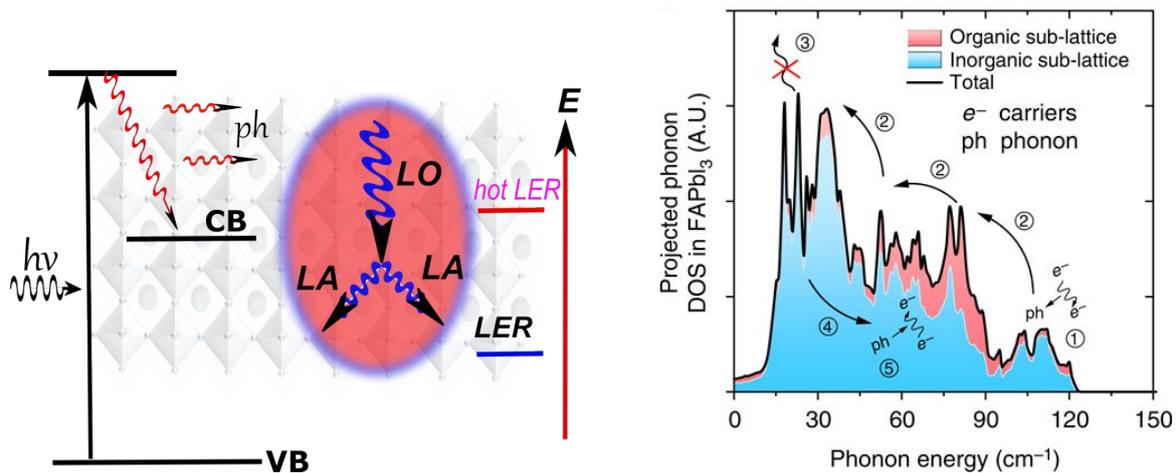

Figure 2. (a) scheme for LER suppress hot carrier cooling. The homojunction with strain of different lattice suppress thermal transport to the surrounding lattice and then induces acoustic-optical phonon upconversion. (b) Proposed phonon dynamics in Perovskites. The labelled phonon dynamic processes are: (1) Fröhlich interaction of carriers primarily on the lead-halide framework; (2) relaxation of lead-halide LO phonon, organic sublattice can be excited by phonon–phonon scattering; (3) propagation of acoustic phonon is blocked due to anharmonic phonon–phonon scatterings; (4) upconversion of acoustic phonons; and (5) carrier reheating.[1]

### 3.2 Carrier recombination and relevant effect

To clearly elucidate the physical impact of LER on the carrier recombinations, it is necessary to emphasize that the meaning of long carrier lifetime includes two aspects of meaning, (1) longer carrier lifetime: in halide perovskites, the observed carrier lifetime is 2-4 orders of magnitude longer than the prediction by Langevin theory,[3, 44-45] spanning from tens of ns to hundreds of µs. And (2) prolonged carrier lifetime under continuous light illumination or excitation,[7-9, 46] as typically expressed as time-dependent PL spectra and time-dependent time-resolved PL (TRPL) in Figure 3, which has been commonly ignored or misunderstood as instability. In time-dependent PL/TRPL measurement, under continuously constant excitation and identical detection conditions, the PL spectra or PL decay curves are consecutively measured. Therefore, the observed variations fully originate from the fluorescence efficiency and are correlated to carrier recombinations. This is generally a reversible process and degradation can be safely excluded.[8-9, 47]

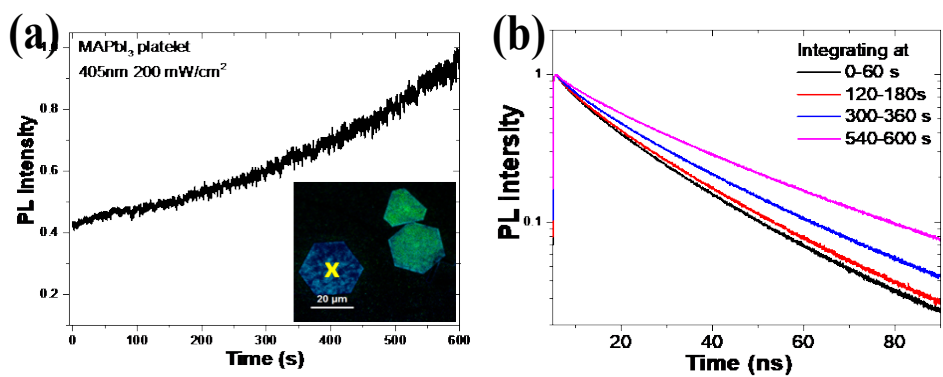

Figure 3 Illumination induced PL enhancement in halide perovskites, ultraslow response in timescale of seconds. PL Intensity (a) and PL decay curves (b) in MAPbI$_3$ nanoplatelet under continuous illumination (405 nm 200 mW/cm$^2$). Inset (a) is the PL imaging of the sample.

Many physical microscopic origins were proposed to interpret such a long lifetime and diffusion length, such as delayed fluorescence (through triplet state) or indirect bandgap transition,[48-53] formation of polarons and thus screening,[32, 54-56] giant Rashba effect,[32, 57-58] ferroelectric domains,[59-61] photon recycling etc.[62-64] These physical origins can partly explain the long carrier lifetime to a certain extent but their rationality is experimentally disproved and thus not extensively accepted.[65-66] Most importantly, these interpretations have ignored the prolonged carrier lifetime under continuous illumination and dynamic defect tolerance.[9-10, 24]

Regarding the origin of delayed fluorescence (type-I thermal activated),[67] it cannot present consistency between the electronic structure (singlet and triplet states), composition and observed carrier lifetime.[68-69] Polaron is considered a quasiparticle that is formed through coulomb interaction between charge carriers (electron or hole) and the surrounding lattice. For the origin of polaron formation, it cannot present a consistent interpretation for the long carrier lifetime and some open questions are hard to address experimentally, such as effective mass, binding energy, coherence length of polarons, the interaction between A-site cation and inorganic sublattice, the ultrashort formation time of polaron in the sub-picosecond timescale, and relationship with detailed compositions.[32, 54, 70-71] In terms of the origin of Rashba band splitting, the spin-forbidden nature of interdomain recombination can reasonably explain the long carrier lifetime in MAPbI$_3$, due to the spin-forbidden indirect transition and thus slow bimolecular recombination.[57, 72] However, nowadays it is difficult to verify the spin-forbidden and spin-allowed recombinations occurring with and without Rashba splitting, respectively, by suitable spectroscopic measurements.[58, 66] Temperature dependent observations of carrier lifetimes contradict the origin of the ferroelectric domain because ferroelectric domains are expected to disappear at high temperatures, which enable faster recombination of photogenerated carriers.[56, 73] The slow carrier lifetime is ascribed to the repeated reabsorption-reemission in the theory of photon recycling.[74] However, there is an apparent problem when interpreting the superior transport properties of hybrid perovskites over the inorganic ones because photon recycling requires the PL quantum yield (PLQY) of the material to be close to unity, but PLQY in MAPbI$_3$ is low.[75-76]

Hot LERs exert a pronounced influence on carrier recombination processes within perovskites, significantly modulating electron-hole recombination. Before hot LER formation, the carrier recombination is similar to that occurring in conventional semiconductors, Figure 4a and 4c. Upon illumination, the photogenerated hot carriers quickly relax to the CB, dominantly by emitting hot phonons.[1, 71] Then the carriers in the CB relax to the valence band by radiative and nonradiative recombinations. The rate equation $\frac{dn}{dt} = \sigma I_{ex}\delta(t) - An - Bn^2 - Cn^3$ has been widely used for describing the carrier recombination dynamics $n = n(t)$ at the CB, taking into account the balanced electron-hole density and free carrier nature in halide perovskites.[77-78] The first term $\sigma I_{ex}\delta(t)$ corresponds to photogenerated carriers that is relevant to the absorption cross section σ and excitation intensity $I_{ex}$, considering this process is ultrafast as a delta function $\delta(t)$. The terms of $An$, $Bn^2$ and $Cn^3$ signify Shockley-Read-Hall (SRH) recombination via subgap trap states, bimolecular recombination, and Auger recombination, respectively.[77]

Upon continuous illumination/excitation, hot LER can form with a higher potential energy than the surrounding lattice. As soon as a subgap carrier moves spatially into the nanodomain of hot LER, it can be efficiently upconverted to the CB; which results in a multiple-circulating recombination process. We can describe as

$$\frac{dn}{dt} = \sigma I_{ex}\delta(t) - An - Bn^2 - Cn^3 + K\sigma' n_{LER} n_{sub} \qquad (1)$$

where $\sigma'$ is the cross section of upconversion from subgap state, $n_{LER}$ and $n_d$ are the densities of hot LER and subgap carriers, and $K$ is a constant. It should be noted that $n_{LER}$ has a very long lifetime (corresponding to extremely low energy dissipation) so the carrier lifetime can be very long, sensitively dependent on the dynamic range of the detection system[3]. At the same time, the net defect trapping will be efficiently decreased and PL efficiency will increase because subgap carrier upconversion functionates actually as a detrapping, which results in defect healing[8-9] and dynamic defect tolerance,[7] Figure 4. It is necessary to note that the upconversion rate depends on the density of hot LER; therefore, the gradual formation of hot LERs leads to (1) a much longer carrier lifetime than that predicted by Langevin theory and (2) a prolonged carrier lifetime with continuous illumination, as dynamic defect tolerance. From an energy point of view, the hot LER provides extra energy to the subgap carrier and results in upconversion. It should note that, the accumulation effect is impossible to avoid in TRPL measurement because experimental acquisition of carrier lifetime cannot be completed in one cycles of excitation-detection, such as fs-ps-ns for transient absorption or ps-ns-us for time-correlated single photon counting (TCSPC). This is also applicable for other transient spectroscopy techniques, which are usually completed in tens of seconds, other than fs-ps-ns-us timescales.

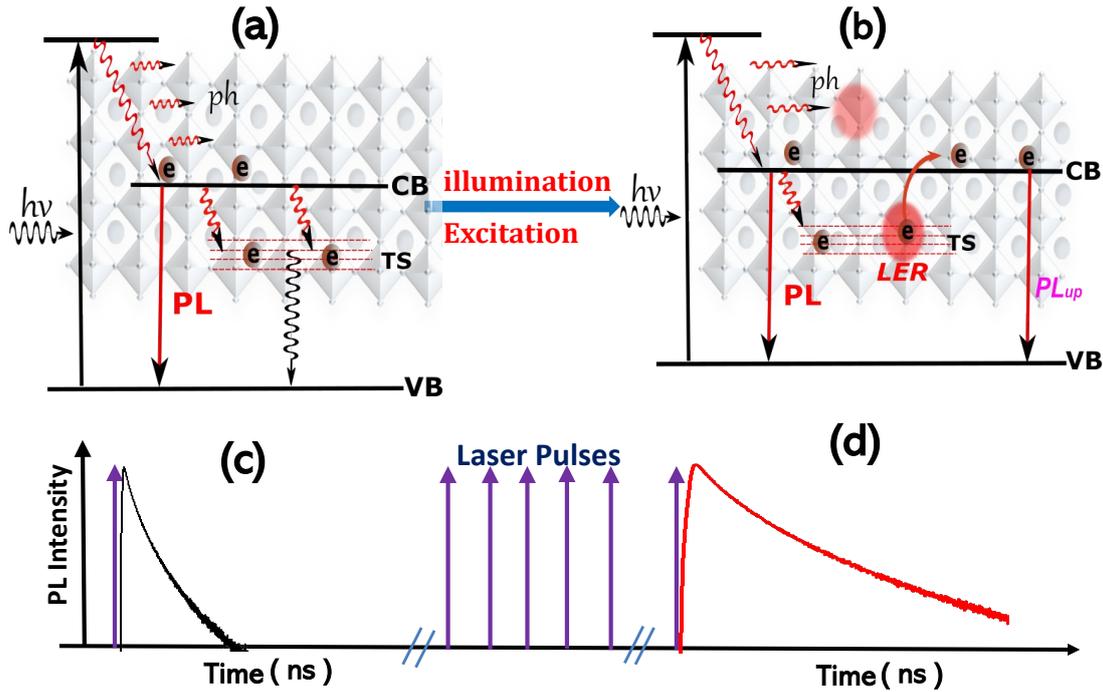

Figure 4. Schematic photobrightening by LER induced subgap carrier upconversion. (a) without LER effect before illumination and (c) display a short carrier lifetime. (b) upon illumination hot LERs form (red regions) and subgap carrier can be upconverted back to the CB, (d) resulting in significantly prolonged carrier lifetime. (TS: trapping/metastable states, ph: phonon, e: electron)

### 3.3. Anomalous ultraslow Phenomena

In addition, hot LERs can give rise to a diverse array of anomalous ultraslow phenomena over extended time scales that defy conventional explanations. As described in the last term of equation 1, the hot LER will result in an increase in the carrier population of the CB, which displays illumination induced PL enhancement (photobrightening), and also dynamic defect tolerance. In time regime, the hot LER induced upconversion of subgap carriers will extend the recombination process through prolonged carrier lifetime. Furthermore, the formation of hot LER can be a gradual accumulation process, that results in a slow increase in PL efficiency and carrier lifetime, as shown in Figure 3. As expressed by the term $K\sigma' n_{LER} n_{sub}$, with continuous excitation, the density of hot LER can gradually increase, and the component of the upconverted subgap carrier increases, which agrees with the observation.

On the other hand, hot LERs facilitate the formation and stabilization of polarons,[79] resulting in additional complexity in the material's electronic and optical behavior. Through their multifaceted impact on perovskite dynamics, hot LERs emerge as key determinants of device performance and functionality, offering insights into the underlying physics governing the exceptional optoelectronic properties of halide perovskite materials. The strain at the surface of LER as a homojunction can suppress thermal transport, therefore LA energy can accumulate in LER and stay for a long time, up to seconds or hours. This feature is key to warranting the observed ultraslow processes, such as persistent structural polarization[11], memristor or memory effect,[12-13, 20] ferroelectric effect[11, 61] and ultraslow variations of PL efficiency and carrier lifetime as dynamic defect tolerance.[9, 21-24]

**3.4 Anomalous upconversion fluorescence**

Recently, efficient anti-Stokes fluorescence (upconversion) has been observed in halide perovskites[5-6, 19], with an energy difference up to 220 meV[70]. It should be noted that the well-known mechanisms of upconversion PL, such as two-photon excitation and lanthanum-doping[80], can be safely excluded. In all-inorganic lead halide perovskite nanocrystals $CsPbA_3$ (A=I, Br, Cl or a mixture of I-Br or Br-Cl), Xiong et al. observed anomalous upconversion fluorescence, that is, band to band fluorescence generated by the excitation of subgap photons[5, 81-82], with an energy gap of up to ~ 8 $K_BT$ ($k_B$ is the Boltzmann constant, $T$ is temperature).[6] Such efficient upconversion fluorescence has been ascribed to resonant multiple-phonon assisted. It is still a question why such a resonant multiple phonon absorption can efficiently occur in materials with relatively high defect densities, because this is a non-coherent process under the competition of the Stokes process.

The upconversion fluorescence can be interpreted by the proposed LER theory, as schematically shown in Figure 5. The incident photons with energy below the bandgap can be partly absorbed because featured subgap absorption has been confirmed in halide perovskdies.[83-84] The energy will dissipate through phonon emission back to the valence band. The active phonons can effectively couple with LERs and result in the formation of hot LERs. The hot LER can upconvert the subgap carriers, which results in efficient upconversion fluorescence. It should be emphasized that an efficient upconversion is expected because this process will be quasi-linear due to the extremely long LER lifetime up to seconds and no coherence/resonance is required. This quasi-linearity has been confirmed by the experiments of excitation fluence dependent fluorescence.[5-6, 19, 82]

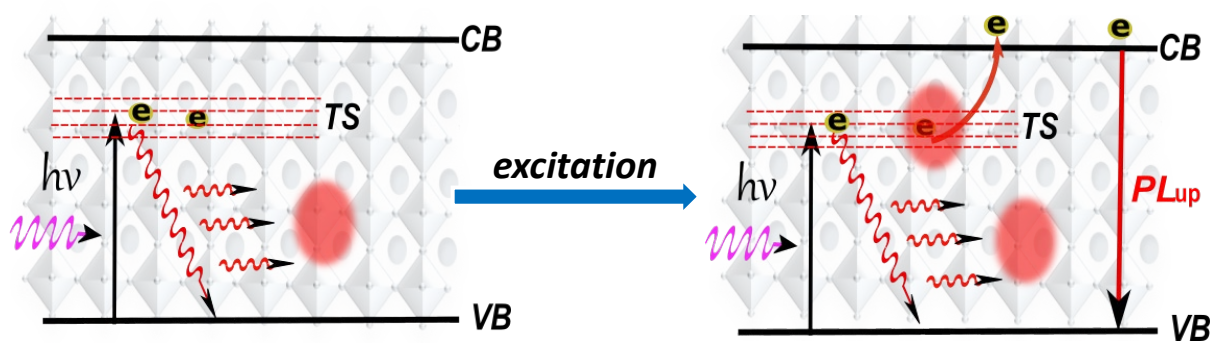

Figure 5. Schematic subgap electron upconversion driven by hot LERs under illumination.

**3.5 Mobile ions**

Mobile ions and various defects are ubiquitous, with a high density for halide perovskites. Mobile ions can form during fabrication or synthesis or are additionally generated post-fabrication by illumination. It is necessary to note that the defect density is quite high for solution-fabricated polycrystal or single crystal perovskites. Mobile ions in halide perovskites mostly include interstitials such as $I^-$, $Br^-$, $MA^+$,

Pb$^+$, FA$^+$ and vacancies such as V$_I$, V$_{MA}$, V$_{Pb}$.[85-87] For mobile ions, the Coulomb effect could be significant to induce lattice distortion and the screening effect of an electric field.[88-89] At the same time, these interstitials and vacancies may act as recombination centers thus impacting electron-hole recombinations, particularly the SRH recombination.[15, 90-91]

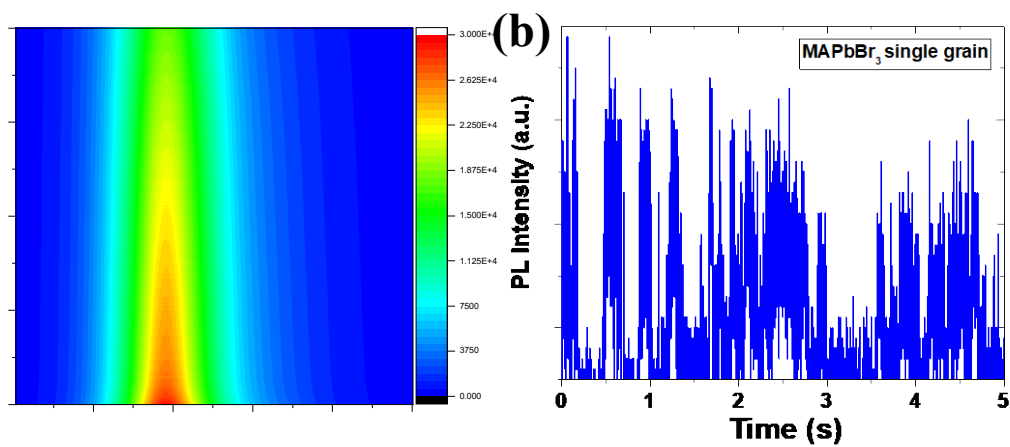

Figure 6 (a) PL spectrum is observed in MAPbBr$_3$ single crystal under 405 nm excitation at a fluence of 350 mW/cm$^2$. (b) Fluorescence intermittency in MAPbBr$_3$ single grain as a function of time under the continuous excitation of 405 nm.

Mobile ions will certainly impact the local lattice and thus indirectly impact LERs. For mobile ions, the charge effect and defect effects will be dominant (mostly negative or detrimental), such as fluorescence intermittency (blinking) in single grain or illumination induced PL quenching in bulk or film,[24, 90, 92] I-V hysteresis,[93-94] and phase segregation.[95-96] Both effects of illumination induced PL quenching and PL enhancement can be observed in the same sample. In other words, PL enhancement is observed at a low excitation fluence; in contrast, a PL quenching (intensity decrease) is observed at higher excitation intensity in the same sample. As an example of MAPbBr$_3$ single crystal, a gradual quenching can be observed under continuous excitation at 350 mW/cm$^2$, Figure 6a; a photobrightening is observed with a fluence below 200 mW/cm$^2$. At a single grain of polycrystal MAPbBr$_3$ film, fluorescence intermittency can be observed (Figure 6b) under continuous excitation at 405 nm.[87, 97-98] The dominance is dependent on the composition, fabrication quality and illumination conditions. This is clearly demonstrated in our recent publication[46].

## 4. Conclusions and outlook

We have elucidated that the LER in halide perovskites can comprehensively impact photogenerated carrier dynamics, from hot carrier cooling to carrier recombination, which mostly resolves the anomalous phenomena observed in halide perovskites, including slowed hot carrier cooling, anomalous anti-Stocks fluorescence and dynamic defect tolerance. We propose that the lattice energy reservoir is the fundamental physical origin for the superior optoelectronic properties and device performance of halide perovskites, given that halide perovskites have the unique features of a soft lattice, subgap absorption and mixed ionic-electronic conduction. Such superior optoelectronic properties of halide perovskites are holistically considered to be the essential cause of the high-performance photovoltaics and optoelectronics.

Our investigation is mostly based on carrier energy relaxation dynamics and discussion from an energy point of view. To date, the LER theory has not been verified using spatial-lattice investigation; therefore, the detailed knowledge is still lacking. It is highly demanded for verification by high resolution electronic microscopy, such as HAADF-STEM, because such investigation is able to provide detailed knowledge for lattice, composition, energy, strain and thermal transport down to the atomic scale, especially for time-energy-correlated TEM and STM. Furthermore, halide perovskites have become a large family, from various compositions, quantum/dielectric confined nanostructures, to double perovskites. In these special circumstances, the LER effect and physical influence are expected to be impacted, with some special variation. Moreover, calculation and simulation for this topic are still lacking because detailed lattice, composite and strain information is not available for LER. The proposed LER theory will provide a framework for energy relaxation for developing theoretical calculations for deep understanding.

## Acknowledgement

Authors acknowledges the financial support from Australian Research Council through Future Fellowship (FT210100806), the Discovery Project (DP220100603), the Centre of Excellence Program (CE230100006) and the Industrial Transformation Training Centre (IC180100005) and from Australian Renewable Energy Agency (ARENA).

**Competing interests**: none declared.